# Decentralized Variational Bayesian UKF with Maximum Generalized Student's t-kernel Correntropy for Wide-Area Power System state estimation

Jinhui Hu, Haiquan Zhao, *Senior Member, IEEE,* and Yi Peng

***Abstract*—A Conventional centralized state estimators exhibit limited robustness in large-scale grids and face practical deployment hurdles. To overcome these challenges, this paper proposes a decentralized maximum generalized Student's t-kernel correntropy Variational Bayesian unscented Kalman filter (D-MGST-VBUKF). The algorithm optimizes the estimation performance at three levels for the regionalized state estimation needs: first, to address non-Gaussian measurement noise in practical systems, we propose the cost function using MGST, retaining Student's t robustness while improving adaptability to complex noise by expanding the degree-of-freedom parameter; secondly, the VB inference framework is constructed to model the unknown noise distribution online, and the joint optimization of the noise statistical characteristics and state estimation is realized by constructing the conjugate prior distribution; finally, the regional state fusion mechanism is established based on the topological correlation characteristics of the power grid, and the global consistency correction of the local estimation results is realized by constructing the state coordination equation of the boundary nodes. Simulation experiments in IEEE 14-bus and IEEE 39-bus system show that the method has stronger robustness compared with the traditional algorithm under non-Gaussian noise environment and unknown noise environment.**

## I. INTRODUCTION

IN recent years, with expanding power system scale and renewable energy integration, traditional state estimation methods relying on local SCADA systems face challenges due to limited scalar measurements and second-level update cycles, which hinder monitoring of remote and weakly-observed nodes in wide-area grids [1]-[5]. While Wide-Area Measurement Systems (WAMS) incorporating Phasor Measurement Units (PMUs) enhance observability through high-dimensional synchronized data [6][7], centralized state estimation frameworks face several critical bottlenecks. These include computational complexity arising from high-dimensional measurements, temporal-spatial synchronization mismatches, and vulnerability to single-point failures [8][9]. To address this, Decentralized State Estimation (DSE) has emerged, decomposing the grid into regions to reduce computational burden and communication bandwidth [7]-[12]. However, a significant gap remains: existing decentralized methods often rely on predefined noise statistics or lack robustness against regional failures, compromising accuracy in large-scale systems where noise characteristics are rarely known priori.

Beyond architectural challenges, real-world measurement data from SCADA and PMU devices are frequently corrupted by non-Gaussian noises arising from sensor anomalies, equipment failures, or external disturbances [4][13]. Traditional Kalman-based methods relying on Gaussian assumptions degrade significantly under such heavy-tailed noise [14]. While information-theoretic criteria like Maximum Correntropy Criterion (MCC) improve robustness, Gaussian kernel-based methods remain sensitive to outliers due to their light-tailed nature [15]. The Student's t-kernel offers heavy-tailed robustness but introduces a critical limitation: parameter coupling. Its single degree-of-freedom parameter governs both kernel shape optimization and noise statistics matching, restricting synergistic optimization and creating performance bottlenecks in complex noise scenarios [16]. Therefore, there is a need for a kernel design that decouples shape adaptation from outlier suppression to handle dynamic noise complexity.

Compounding the noise distribution issue is the challenge of unknown noise statistics. Power system noise exhibits time-varying, non-stationary statistics due to electromagnetic interference and environmental disturbances. Deviations between assumed and actual noise severely degrade algorithm performance. While adaptive methods like Sage-Husa [17] and Variational Bayesian Kalman filters (VBKF) [18]-[23] attempt online noise estimation, they often oversimplify noise characteristics (neglecting heavy tails) or rely on Gaussian likelihoods which fail under non-Gaussian conditions. Consequently, a framework is required that can jointly optimize kernel function properties and noise statistics matching without relying on predefined distributions.

To implement this framework, the choice of filtering foundation is critical. Among nonlinear methods, the Unscented Kalman Filter (UKF) [24][25][26] is selected over the Extended Kalman Filter (EKF) [27][28] to avoid first-order linearization errors and Jacobian calculations. Compared to the Cubature Kalman Filter (CKF) [29][30], the UKF provides additional flexibility through its sigma-point spread parameter, allowing independent optimization for each region in a decentralized setting. Furthermore, unlike the Ensemble

This work was partially supported by National Natural Science Foundation of China (grant: 62171388, 61871461, 61571374). (Corresponding author: Haiquan Zhao).

Jinhui Hu (e-mail: jhhu_swjtu@126.com), Haiquan Zhao (e-mail: hqzhao_swjtu@126.com) and Yi Peng (e-mail: pengyi1007@163.com) are with the Key Laboratory of Magnetic Suspension Technology and Maglev Vehicle, Ministry of Education, School of Electrical Engineering Southwest Jiaotong University Chengdu, China.

Kalman Filter (EnKF) [31], the UKF's deterministic sigma points avoid sampling noise, which is essential for the stability of iterative variational Bayesian inference. Thus, the UKF provides the optimal computationally efficient foundation for integrating robust kernel methods and VB adaptation. While other robust filtering techniques exist, they present specific challenges for wide-area decentralized estimation. Particle Filters (PF) [32], though capable of handling non-Gaussian noise, suffer from high computational complexity and particle degeneracy in high-dimensional state spaces, limiting their scalability for large-scale grids. H-infinity filters (HIF), although robust, tend to be overly conservative by minimizing worst-case errors, potentially compromising accuracy under normal operating conditions [33]. In contrast, the UKF strikes a balanced trade-off between computational efficiency and the ability to handle nonlinearities. By enhancing the UKF with the proposed MGST kernel and variational Bayesian inference, we achieve robustness against non-Gaussian noise comparable to that of PF, while maintaining the computational scalability required for decentralized wide-area systems.

Therefore, we summarize the key challenges addressed in this work as follows: 1) Gaussian kernel limitations: Light-tailed characteristics and fixed bandwidth make it sensitive to heavy-tailed noise, causing mismatches between kernel shape optimization and noise suppression demands, leading to accuracy loss; 2) Unknown noise challenges: Traditional methods rely on known noise distributions, rarely available in practice, resulting in performance degradation under dynamic and unknown conditions. Emerging solutions like Student's t-distribution modeling and maximum correntropy-based filters show promise but face challenges such as parameter coupling or computational complexity. To overcome these, this paper proposes a decentralized variational Bayesian UKF with the maximum generalized Student's t-kernel correntropy(D-MGST-VBUKF). Key contributions include:

1) A modified generalized Student's t-kernel is presented, which effectively solves the problems of the Gaussian kernel's difficulty in coping with the heavy-tailed noise and the insufficient degree of freedom of the Student's t-kernel's parameters;

2) A robust variational Bayesian framework for local state estimation is constructed to achieve the estimation of local noise distribution in wide-area power systems.

3) A novel D-MGST-VBUKF algorithm is proposed to address the critical challenge of wide-area power system state estimation, demonstrating superior performance in achieving accurate state assessments under conditions of both unknown noise characteristics and complex non-Gaussian noise distributions.

The rest of the paper is organized as follows: the decentralized power system model is introduced in Section II, while the proposed MGST and the flow of the proposed algorithm are given in Section III, the simulation results are given in Section IV, and finally, the conclusion is given in Section V.

## II. Power System Model

### A. Power system state transfer model

Consider a power system model containing $\boldsymbol{K}$ buses with a quasi-steady state transfer equation for the *k*-th bus [4][7]

$$\mathbf{v}_{k,m} = \mathbf{F}_k \mathbf{v}_{k,m-1} + \mathbf{G}_k \bar{\mathbf{v}}_k + \mathbf{q}_{k,m} \tag{1}$$

where $\mathbf{F}_k \in \mathbb{R}^{2\times 2}$ is a known matrix of reaction state change rates, $\bar{\mathbf{v}}_k$ is the desired steady state; $\mathbf{G}_k = (\mathbf{I} - \mathbf{F}_k) \in \mathbb{R}^{2\times 2}$ used to explain trends in state trajectories. $\mathbf{v}_{k,m} = \left[U_{k,m}, \theta_{k,m}\right]^T \in \mathbb{R}^{2\times 1}$ denotes the state vector of the *k*-th bus at the *m*-th moment, where $U_{k,m}, \theta_{k,m}$ represent the voltage magnitude and voltage phase angle of the *k*-th bus at *m*-th moment, respectively; $\mathbf{q}_{k,m}$ is the Gaussian white noise with zero-mean and covariance matrix $\mathbf{Q}_{k,m}$.

### B. SCADA and PMU Measurement model

A hybrid measurement model of SCADA and PMU is used to make observations of the state in the power system. SCADA units are installed across all buses to measure bus voltage magnitudes and power flows between interconnected buses. In contrast, PMUs, deployed on a subset of buses due to their higher implementation cost compared to SCADA, measure bus voltage phasors.

At the *m*-th moment, the SCADA unit provides the voltage magnitude at the *k*-th bus [7]

$$\mathbf{z}^s_{k,m} = \mathbf{H}_s \mathbf{v}_{k,m} + \mathbf{r}^s_{k,m} \tag{2}$$

where $\mathbf{H}_s = [1 \quad 0]$ ; $\mathbf{r}^s_{k,m}$ is the noise of the SCADA measurements, which satisfies the properties of a Gaussian distribution with zero-mean and a covariance matrix of $\mathbf{R}^s_{k,m}$. Then the active and reactive power flows between buses k, j at moment m measured by SCADA are given by

$$\mathbf{z}^s_{kj,m} = \mathbf{h}^s_{kj,m}\left(\mathbf{v}_{k,m}, \mathbf{v}_{j,m}\right) + \mathbf{r}^s_{kj,m} \tag{3}$$

where $\mathbf{z}^s_{kj,m} = \left[P_{kj,m} \quad Q_{kj,m}\right]$, $\mathbf{r}^s_{kj,m}$ is also the noise of the SCADA measurements, which satisfies the properties of a Gaussian distribution with a zeros-mean and a covariance matrix of $\mathbf{R}^s_{kj,m}$. And $\mathbf{h}^s_{kj,m}(\bullet)$ is given by [4][7][13]

$$\begin{aligned} P_{kj,m} &= \mathbf{U}^2_{k,m}\left(g_{sk} + g_{kj}\right) - \\ &\mathbf{U}_{k,m}\mathbf{U}_{j,m}\left(g_{si}\cos\left(\theta_{k,m} - \theta_{j,m}\right) + b_{si}\cos\left(\theta_{k,m} - \theta_{j,m}\right)\right) \end{aligned} \tag{4}$$

$$\begin{aligned} Q_{kj,m} &= -\mathbf{U}^2_{k,m}\left(b_{sk} + b_{kj}\right) - \\ &\mathbf{U}_{k,m}\mathbf{U}_{j,m}\left(g_{si}\sin\left(\theta_{k,m} - \theta_{j,m}\right) - b_{si}\cos\left(\theta_{k,m} - \theta_{j,m}\right)\right) \end{aligned} \tag{5}$$

where $g_{sk} + jb_{sk}$ represents the conductance and electrical energy rate of the shunt at *k*-th bus, and $g_{kj} + jb_{kj}$ represents the conductance and electrical energy rate between *k*-th bus and *j*-th bus.

Then the measurement model of the voltage phase measured by the PMU is given by [7]

$$\mathbf{z}^p_{k,m} = \mathbf{H}_p \mathbf{v}_{k,m} + \mathbf{r}^p_{k,m} \tag{6}$$

where $\mathbf{H}_p$ is a unit matrix, and $\mathbf{r}_{k,m}^{p}$ is the noise of the PMU measurements with a Gaussian distribution, mean 0, and covariance matrix $\mathbf{R}_{k,m}^{p}$.

### *C. Decentralized system model*

In order to obtain a more refined local state estimate of the power system, we divide the system into $N$ regions. The system model for each region and the edge measurement model between two regions are respectively given as follows.

First, the local system model gives the state transfer model for the $n$-th region [7]

$$\mathbf{v}_{n,m} = \overline{\mathbf{F}}_n \mathbf{v}_{n,m-1} + \overline{\mathbf{G}}_n \overline{\mathbf{v}}_n + \overline{\mathbf{q}}_{n,m} \tag{7}$$

where $\mathbf{v}_{n,m} = \left[ \mathbf{v}_{n_1,m}^{T} \quad \mathbf{v}_{n_2,m}^{T} \quad \cdots \quad \mathbf{v}_{n_{\tau_n},m}^{T} \right]^{T} \in \mathbb{R}^{\alpha\times 1}$ is the state vector of $n$-th region at $m$-th moment; $\overline{\mathbf{F}}_n = \left\{ \overline{\mathbf{F}}_{n_1} \quad \overline{\mathbf{F}}_{n_2} \quad \cdots \quad \overline{\mathbf{F}}_{n_{\tau_n}} \right\}$ is the transfer matrix of $n$-th region at $m$-th moment; $\overline{\mathbf{G}}_n = I - \overline{\mathbf{F}}_n$ is related to the trend behavior of the state estimates for $n$-th region; $\overline{\mathbf{v}}_n = \left[ \overline{\mathbf{v}}_{n_1}^{T} \quad \overline{\mathbf{v}}_{n_2}^{T} \quad \cdots \quad \overline{\mathbf{v}}_{n_{\tau_n}}^{T} \right]^{T}$ then the expected steady state of $n$-th region; $\overline{\mathbf{q}}_{n,m} = \left[ \overline{\mathbf{q}}_{n_1,m}^{T} \quad \overline{\mathbf{q}}_{n_2,m}^{T} \quad \cdots \quad \overline{\mathbf{q}}_{n_{\tau_m},m}^{T} \right]^{T}$ then the process noise. $\tau_n$ denotes the total number of buses in $n$-th region.

Then the local measurement model of $n$-th region is given by

$$\mathbf{z}_{n,m} = \mathbf{h}_n \left( \mathbf{v}_{n,m} \right) + \mathbf{r}_{n,m} \tag{8}$$

where

$$\mathbf{z}_{n,m} = \begin{bmatrix} \left(\mathbf{z}_{n_1,m}^{s}\right)^{T}, \ \ldots, \left(\mathbf{z}_{n_{\tau_n},m}^{s}\right)^{T}, \\ \ldots, \ \left(\mathbf{z}_{n_a n_c,m}^{s}\right)^{T}, \ \ldots, \ \left(\mathbf{z}_{n_b,m}^{p}\right)^{T}, \ \ldots \end{bmatrix}^{T} \in \mathbb{R}^{\beta\times 1} \tag{9}$$

$$\mathbf{r}_{n,m} = \begin{bmatrix} \left(\mathbf{r}_{n_1,m}^{s}\right)^{T}, \ \ldots, \left(\mathbf{r}_{n_{\tau_n},m}^{s}\right)^{T}, \\ \ldots, \ \left(\mathbf{r}_{n_a n_c,m}^{s}\right)^{T}, \ \ldots, \ \left(\mathbf{r}_{n_b,m}^{p}\right)^{T}, \ \ldots \end{bmatrix}^{T} \in \mathbb{R}^{\beta\times 1} \tag{10}$$

Here, the calculation of $\mathbf{h}_n(\bullet)$ contains (2), (3) and (6), and the measurements from SCADA and PMU are aggregated to obtain a localized edge measurement model of the measurements.

Since some of the buses in neighboring areas are connected to each other, the measurements in this part are defined as the edge measurement model, which for area a can be given as

$$\mathbf{z}_{ab,m} = \mathbf{h}_{ab} \left( \mathbf{v}_{a,m}, \mathbf{v}_{b,m} \right) + \mathbf{r}_{ab,m} \tag{11}$$

where $\mathbf{h}_{ab}(\bullet)$ is calculated in the same way as (3); $\mathbf{r}_{ab,m}$ denotes the measurement noise of the edge measurement model, which satisfies a Gaussian distribution with mean 0 and covariance matrix $\mathbf{R}_{ab,m}^{s}$.

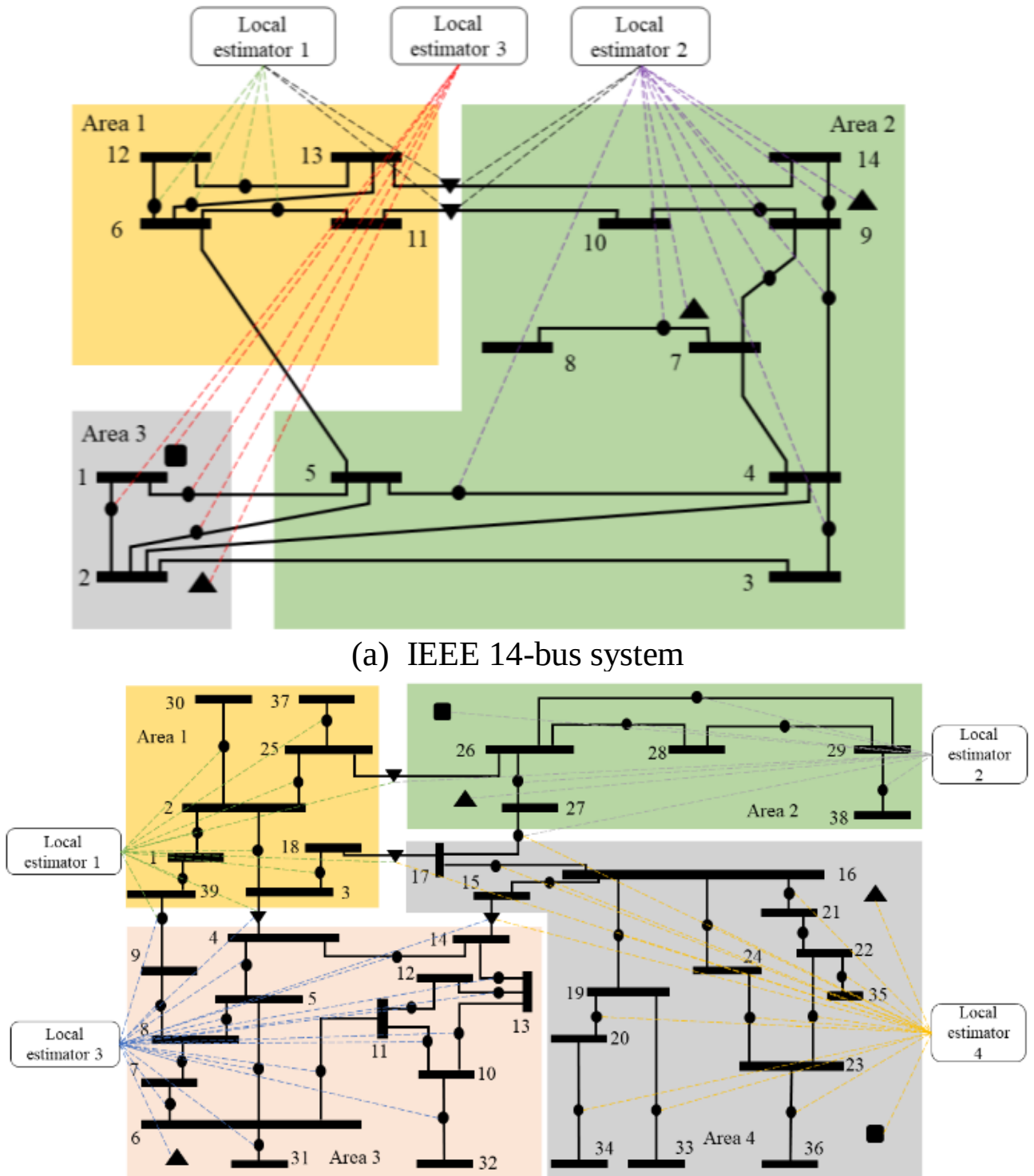

Fig. 1. Schematic diagram of decentralized partitioning of wide-area power system

***Remark 1:*** In real power systems, measurements are often subject to a variety of anomalous data disturbances, and such anomalous disturbances are often unknown. We model these disturbances in the form of noise in the measurement model, and traditional state estimation methods often require that the distribution of the various types of noise is known. As a result, the performance of such methods degrades in the presence of unknown anomalous disturbances, which requires us to improve the robustness of the algorithm by using methods that can estimate the noise distribution.

***Remark 2:*** Power system measurement models are subject to disturbances from outliers, which tend to exhibit non-Gaussian properties in state estimators. And traditional estimators are often based on Gaussian assumptions when dealing with noise disturbances. Although there have been some methods that use information theoretic learning to realize robust state estimation, all of them have different degrees of shortcomings. Such as the degradation of performance in complex environments, the problem of higher computational complexity of the algorithms, etc. Therefore, more advanced robust methods are needed to realize state estimation of power systems.

***Remark 3:*** It is important to note that while the state transition model in (1) is linear, the overall system is not fully linear due to the nonlinear measurement model $\mathbf{h}(\cdot)$ described in (2)-(6). The power flow measurements (4) involve trigonometric functions and products of state variables, introducing significant nonlinearities. Therefore, a standard linear KF is not applicable. Among nonlinear filters, UKF is preferred: it avoids EKF's linearization errors and Jacobian calculations; offers greater tuning flexibility than CKF via parameter λ, facilitating VB integration; and unlike EnKF, its deterministic sigma points are computationally efficient and free of sampling noise, which is critical for stable VB inference. Its effectiveness is well-established in power system state estimation [12]-[14], making it a solid foundation for our decentralized robust framework. Therefore, the UKF provides the optimal foundation for integrating the proposed MGST kernel and VB adaptation in a computationally efficient and theoretically sound manner.

## III. Algorithms

This section details the proposed Decentralized Maximum Generalized Student's t-kernel Correntropy Variational Bayesian UKF (D-MGST-VBUKF). Distinct from existing decentralized robust UKF approaches [7] which typically assume known noise statistics, and standard Variational Bayesian Kalman Filters (VBKF) [31]-[36] which rely on Gaussian likelihoods, our framework introduces three key innovations:

Unlike standard Student's t-kernels where shape and tail decay are coupled, the proposed MGST kernel decouples these via the shape parameter $\xi$, allowing independent adaptation to kurtosis and outliers.

While traditional robust filters require offline noise tuning, our VB framework online estimates the inverse Wishart parameters for both process and measurement noise, addressing the unknown noise challenge.

We extend the local VB-MGST estimator into a decentralized architecture using boundary node coordination, ensuring global consistency without a central fusion center.

The following subsections elaborate on the MGST criterion, the local VB-UKF derivation, and the decentralized fusion mechanism.

To improve readability, we simplify the notation for the state estimation process. Table I summarizes the key notations used in this section. Specifically, we denote the prior state estimate at time $m$ for region $n$ as $\mathbf{v}_{n,m}^{-}$ and the posterior estimate as $\mathbf{v}_{n,m}$. Similarly, $\mathbf{P}_{n,m}^{-}$ and $\mathbf{P}_{n,m}$ represent the prior and posterior error covariance matrices, respectively. The measurement noise covariance is denoted as $\mathbf{R}_{n,m}$.

Table I
Key Notations

| Symbol | Description |
|---|---|
| $n$, $m$ | Region index, Time step index |
| $\mathbf{v}_{n,m}^{-}$, $\mathbf{v}_{n,m}$ | Prior and Posterior state vector |
| $\mathbf{P}_{n,m}^{-}$, $\mathbf{P}_{n,m}$ | Prior and Posterior error covariance |
| $\mathbf{R}_{n,m}$ | Measurement noise covariance |
| $\mathbf{z}_{n,m}$ | Measurement vector |
| $\xi, c$ | Shape and degrees-of-freedom parameters |
| $j$ | Variational Bayesian iteration index |
| $i$ | Sigma point index |

### *A. Maximum generalized Student's t-kernel correntropy criterion*

Conventional correntropy criteria often utilize Gaussian kernels, which suffer from sensitivity to outliers due to their light-tailed nature. While the Student's t-kernel improves robustness, its single degree-of-freedom parameter couples tail decay with kernel shape, limiting adaptability. To address this, we employ the Maximum Generalized Student's t (MGST) kernel defined as:

$$S_{c,\gamma}(\mathbf{v}) = \left(1 + \frac{\|\mathbf{v}\|^{\xi}}{c\gamma^{\xi}}\right)^{-(c+\xi)/\xi} \tag{12}$$

where $\mathbf{v}$ is the error vector, $c$ controls the tail decay (robustness to outliers), and $\xi$ governs the kernel curvature (adaptability to kurtosis).

***Remark 3:*** The introduction of $\xi$ allows the kernel to independently adjust to the peakedness of the noise distribution without altering its outlier suppression capability governed by $c$. This decoupling is critical for power system states where noise characteristics vary dynamically between steady-state (low kurtosis) and transient events (high kurtosis). Fig. 2 illustrates how varying $\xi$ adjusts the kernel shape while maintaining heavy-tailed properties in different error size v.

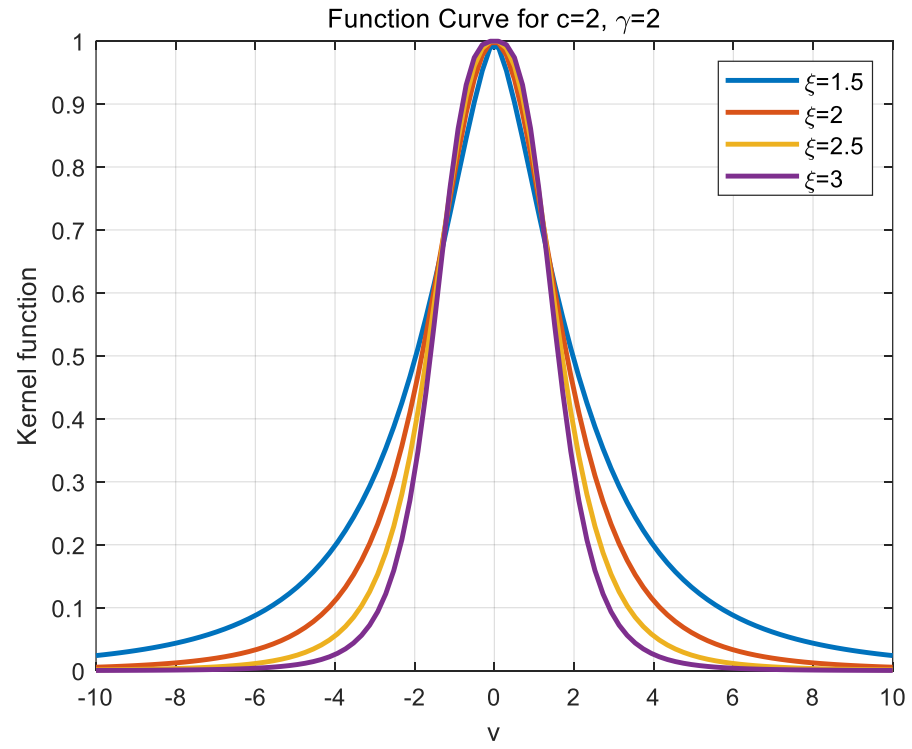

Fig. 2. Comparison of the parameter $\xi$ for different values of the kernel function shape.

### *B. Local MGST Variational Bayesian UKF*

The local estimator integrates the UKF for state propagation with VB inference for noise covariance estimation. The process is divided into two phases:

***1) VB-Enhanced Prediction:***

Consider first the local state estimate of $n$-th region. Here, the priori state estimate $\mathbf{v}_{n,m-1}^{-}$ and the priori error covariance matrix $\mathbf{P}_{n,m-1}^{-}$ of the algorithm at $m$-th moment are computed using the unscented transformation. Therefore, we first assume that the posteriori state estimate $\mathbf{v}_{n,m-1}$ as well as the posteriori error covariance matrix $\mathbf{P}_{n,m-1|m-1}$ at $m$-1-th moment are known. Then we use the unscented transformation to obtain the sigma points:

$$v_{n,m-1}^{i} = \begin{cases} \mathbf{v}_{n,m-1} \\ \mathbf{v}_{n,m-1} + \left(\sqrt{\alpha+\varphi}\mathbf{S}_{n,m-1}\right)_i, i = 1,\dots\alpha \\ \mathbf{v}_{n,m-1} - \left(\sqrt{\alpha+\varphi}\mathbf{S}_{n,m-1}\right)_{i-\alpha}, i = \alpha+1,\dots 2\alpha \end{cases} \tag{13}$$

where $\mathbf{S}_{n,m-1}\mathbf{S}_{n,m-1}^{T} = \mathbf{P}_{n,m-1}$; $\varphi$ is a proportionality correction coefficient that takes the value of $\varphi = \lambda^2(\alpha+\eta) - \alpha$, where $0 < \lambda < 1$ determines the distribution of sigma points, and $\eta$ is used to mitigates truncation errors in the mean-covariance propagation for high-dimensional state spaces. The obtained sigma points are then used to calculate $\bar{\mathbf{v}}_{n,m}$ and $\mathbf{P}^{-}_{n,m}$:

$$\bar{\mathbf{v}}_{n,m}=\sum_{0}^{2\alpha}\omega_v^i\mathbf{f}\left(\nu^i_{n,m-1}\right) \tag{14}$$

$$\begin{aligned}\mathbf{P}^{-}_{n,m}=&\sum_{0}^{2\alpha}\omega_c^i\left[\mathbf{f}\left(\nu^i_{n,m-1}\right)-\bar{\mathbf{v}}_{n,m}\right]\\&\times\left[\mathbf{f}\left(\nu^i_{n,m-1}\right)-\bar{\mathbf{v}}_{n,m}\right]^T+\mathbf{Q}_{n,m-1}\end{aligned} \tag{15}$$

where $\mathbf{f}(\bullet)$ is state transfer function, which can be obtained by (1). And

$$\begin{cases}\omega_v^0=\dfrac{\varphi}{\alpha+\varphi}\\ \omega_c^0=\dfrac{\varphi}{\alpha+\varphi}+1+\tau-\lambda^2\\ \omega_v^i=\omega_c^i=\dfrac{1}{2(\alpha+\varphi)}\end{cases} \tag{16}$$

It should be noted here that $\mathbf{P}^{-}_{n,m}$ obtained from the calculation in (1) is inaccurate due to the fact that the state noise covariance matrices $\mathbf{Q}^{-}_{n,m}$ and $\mathbf{R}_{n,m}$ in the real power system are unknown and $\mathbf{P}_{n,m-1}$ calculated at the moment *m*-1 is inaccurate. Therefore, we need to perform further calculations for $\mathbf{P}^{-}_{n,m}$ and $\mathbf{R}_{n,m}$. For Gaussian distributions with uncertain variance but known means, the Inverse-Wishart (IW) distribution is a method commonly used to characterize their covariance prior distributions. For an unknown noise distribution $\mathbf{R}_{n,m}$ and covariance matrix $\mathbf{P}^{-}_{n,m}$, the IW distribution is modeled with the following PDF [18]:

$$\begin{aligned}&\rho\left(\mathbf{P}^{-}_{n,m}\mid\mathbf{z}_{n,1:m-1}\right)=\mathrm{IW}\left(\mathbf{P}^{-}_{n,m};\bar{\delta}_{n,m};\bar{\Delta}_{n,m}\right)\\&=\frac{\left(\left|\bar{\Delta}_{n,m}\right|\big/2^{\theta}\right)^{t_k/2}}{\Gamma_\theta\left(\bar{\delta}_{n,m}\big/2\right)}\bullet\frac{\exp\left(-0.5tr\left(\bar{\Delta}_{n,m}\left(\mathbf{P}^{-}_{n,m}\right)^{-1}\right)\right)}{\left|\mathbf{P}^{-}_{n,m}\right|^{\left(\bar{\delta}_{n,m}+\theta+1\right)/2}};\bar{\delta}_{n,m}>\theta-1\\&\rho\left(\mathbf{R}_{n,m}\mid\mathbf{z}_{n,1:m-1}\right)=\mathrm{IW}\left(\mathbf{R}_{n,m};\hat{\bar{\imath}}_{n,m};\hat{\bar{l}}_{n,m}\right)\\&=\frac{\left(\left|\hat{\bar{l}}_{n,m}\right|\big/2^{\theta}\right)^{t_k/2}}{\Gamma_\theta\left(\hat{\bar{\imath}}_{n,m}\big/2\right)}\bullet\frac{\exp\left(-0.5tr\left(\hat{\bar{l}}_{n,m}\mathbf{R}^{-1}_{n,m}\right)\right)}{\left|\mathbf{R}_m\right|^{\left(\hat{\bar{\imath}}_{n,m}+\theta+1\right)/2}};\hat{\bar{\imath}}_{n,m}>\theta-1\end{aligned} \tag{17, 18}$$

where $\hat{\bar{\imath}}_{n,m},\hat{\bar{l}}_{n,m}$ represent the coefficients of freedom and the inverse scale matrix of $\rho\left(\mathbf{R}_{n,m}\mid\mathbf{z}_{n,1:m-1}\right)$, and $\bar{\delta}_{n,m},\bar{\Delta}_{n,m}$ represent the coefficients of freedom and the inverse scale matrix of $\rho\left(\mathbf{P}_{n,m|m-1}\mid\mathbf{z}_{n,1:m-1}\right)$, respectively. Next determine the priori parameters $\hat{\bar{\imath}}_{n,m},\hat{\bar{l}}_{n,m},\bar{\delta}_{n,m}$ and $\bar{\Delta}_{n,m}$. First, in order to obtain the priori information about $\mathbf{P}^{-}_{n,m}$, set the mean of $\mathbf{P}^{-}_{n,m}$ to the rated $\bar{\mathbf{P}}_{n,m}$.

$$\bar{\mathbf{P}}_{n,m}=\frac{\bar{\Delta}_{n,m}}{\bar{\delta}_{n,m}-\alpha-1}=\mathbf{P}^{-}_{n,m}-\mathbf{Q}_{n,m-1}+\mathbf{Q}_{n,m-1} \tag{19}$$

where $\mathbf{Q}_{n,m-1}$ denotes the parameter of rated $\bar{\mathbf{P}}_{n,m}$. Then make

$$\bar{\delta}_{n,m}=\alpha+\varsigma+1 \tag{20}$$

where $0<\varsigma<1$ is the tuning parameter. Carrying (20) into (19) yields

$$\bar{\Delta}_{n,m}=\varsigma\bar{\mathbf{P}}_{n,m} \tag{21}$$

Next, the prior is computed for the noise covariance matrix $\mathbf{R}_{n,m}$. $\mathbf{R}_{n,m}$ is unknown and time-varying, and first we give its prior distribution according to variational Bayesian theory:

$$\rho\left(\mathbf{R}_{n,m}\mid\mathbf{z}_{n,1:m-1}\right)=\int\rho\left(\mathbf{R}_{n,m}\mid\mathbf{R}_{n,m-1}\right)\rho\left(\mathbf{R}_{n,m-1}\mid\mathbf{z}_{n,1:m-1}\right)d\mathbf{R}_{n,m-1} \tag{22}$$

where $\rho\left(\mathbf{R}_{n,m-1}\mid\mathbf{z}_{n,1:m-1}\right)$ is the posterior probability density function of $\mathbf{R}_{n,m-1}$. Based on the recurrence relation, we get

$$\rho\left(\mathbf{R}_{n,m-1}\mid\mathbf{z}_{n,1:m-1}\right)=\mathrm{IW}\left(\mathbf{R}_{n,m-1};\hat{\imath}_{n,m-1};\hat{l}_{n,m-1}\right) \tag{23}$$

Next, considering the slowly varying noise distribution in the power system, the following equation is used to calculate the a priori parameters $\hat{\bar{\imath}}_{n,m}$ and $\hat{\bar{l}}_{n,m}$ obtained via

$$\hat{\bar{\imath}}_{n,m}=\zeta\left(\hat{\bar{\imath}}_{n,m}-\beta-1\right)+\beta+1 \tag{24}$$

$$\hat{\bar{l}}_{n,m}=\zeta\hat{l}_{n,m-1} \tag{25}$$

where $\zeta$ is the forgetting factor.

We then proceeded to predict the measurements using the traceless transform, generating Sigma points using the priori state estimation $\mathbf{v}_{n,m|m-1}$

$$\nu^i_{n,m|m-1}=\begin{cases}\bar{\mathbf{v}}_{n,m}\\ \bar{\mathbf{v}}_{n,m}+\left(\sqrt{\alpha+\varphi}\mathbf{S}^{-}_{n,m}\right)_i,i=1,\ldots\alpha\\ \bar{\mathbf{v}}_{n,m}-\left(\sqrt{\alpha+\varphi}\mathbf{S}^{-}_{n,m}\right)_{i-\alpha},i=\alpha+1,\ldots2\alpha\end{cases} \tag{26}$$

Using the generated sigma points to calculate the predicted measurements as well as the mutual covariance matrix:

$$\hat{\bar{\mathbf{z}}}_{n,m}=\frac{1}{2\alpha+1}\sum_{0}^{2\alpha}\omega_v^i\mathbf{h}\left(\nu^i_{n,m|m-1}\right) \tag{27}$$

$$\mathbf{P}^{-}_{n,\mathbf{vz},m}=\sum_{0}^{2\alpha}\omega_c^i\left[\mathbf{h}\left(\nu^i_{n,m|m-1}\right)-\hat{\bar{\mathbf{z}}}_{n,m}\right]\left[\nu^i_{n,m|m-1}-\bar{\mathbf{v}}_{n,m}\right]^T \tag{28}$$

$$\mathbf{P}^{-}_{n,\mathbf{zz},m}=\sum_{0}^{2\alpha}\omega_c^i\left[\mathbf{h}\left(\nu^i_{n,m|m-1}\right)-\hat{\bar{\mathbf{z}}}_{n,m}\right]\left[\mathbf{h}\left(\nu^i_{n,m|m-1}\right)-\hat{\bar{\mathbf{z}}}_{n,m}\right]^T \tag{29}$$

***Remark 4:*** The $\mathbf{R}_{n,m}$ computed in the priori state estimation is the overall $\mathbf{R}_{n,m}$ of the local state estimation, which includes information about the $\mathbf{R}^p_{n,m}$ of the PMU and $\mathbf{R}^s_{n,m}$ of SCADA measurement data. We identify it here by using $\mathbf{R}_{n,m}$ for uniform statistics, and we do not estimate it here for the $\mathbf{R}^s_{ab,m}$ of the edge measurement model.

*2) Iterative VB Measurement Update:*

The VB inference proceeds in three logical steps:

Factorization: Assume the posterior distribution factorizes into state, process noise covariance, and measurement noise covariance.

Optimization: Minimize the Kullback-Leibler divergence (KLD) between the true posterior and the factorized approximation .

Fixed-Point Iteration: Update the parameters of the Inverse-Wishart distributions for noise covariances and the Gaussian distribution for the state iteratively until convergence.

Then, explicitly state what equations correspond to which step.

To compute the joint distributions of $\bar{\mathbf{v}}_{n,m}$, $\mathbf{P}_{n,m}^{-}$ and $\mathbf{R}_{n,m}$, we employ a variational Bayesian approach. We approximate the true posterior distribution $\rho(\bullet)$ with a factorized distribution $q(\bullet)$ using the mean-field assumption:

$$\rho\left(\mathbf{v}_{n,m},\mathbf{P}_{n,m}^{-},\mathbf{R}_{n,m} \mid \mathbf{z}_{n,1:m}\right)=q\left(\mathbf{v}_{n,m}\right)q\left(\mathbf{P}_{n,m}^{-}\right)q\left(\mathbf{R}_{n,m}\right) \tag{30}$$

where $q(\bullet)$ is the unknown approximate posterior distribution. One can compute $q\left(\mathbf{v}_{n,m}\right)$ , $q\left(\mathbf{R}_{n,m}\right)$ and $q\left(\mathbf{P}_{n,m}^{-}\right)$ by minimizing the Kullback-Leibler divergence (KLD):

$$\begin{aligned}&\left\{q\left(\mathbf{v}_{m}\right),q\left(\mathbf{P}_{m}^{-}\right),q\left(\mathbf{R}_{m}\right)\right\}=\arg\min \mathrm{KLD}\\&\left(q\left(\mathbf{v}_{m}\right)q\left(\mathbf{P}_{m}^{-}\right)q\left(\mathbf{R}_{m}\right)\|\,\rho\left(\mathbf{v}_{m},\mathbf{P}_{m}^{-},\mathbf{R}_{m} \mid \mathbf{z}_{1:m}\right)\right)\end{aligned} \tag{31}$$

The above equation is then solved to obtain

$$\log q(\psi)=E_{\Theta_m^{-\psi}}\left[\log\rho\left(\Theta_m,\mathbf{z}_{1:m}\right)\right]+\sigma_{\psi} \tag{32}$$

$$\Theta_m \triangleq \left[\mathbf{v}_m,\mathbf{P}_m^{-},\mathbf{R}_m\right] \tag{33}$$

where $\Theta_m^{-\psi}$ denotes all the members of $\Theta_m$ whose expectation is $\psi$ and $\sigma_{\psi}$ is the constant associated with $\psi$. Through compositional propagation of probability and predictive distributions, the cascade distribution can be computed by

$$\begin{aligned}\rho\left(\Theta_m,\mathbf{z}_{1:m}\right)&=\rho\left(\mathbf{z}_m \mid \mathbf{v}_m,\mathbf{R}_m\right)\rho\left(\mathbf{v}_m \mid \mathbf{z}_{1:m-1},\mathbf{P}_m^{-}\right)\\&\times\rho\left(\mathbf{P}_m^{-} \mid \mathbf{z}_{1:m-1}\right)\rho\left(\mathbf{R}_m \mid \mathbf{z}_{1:m-1}\right)\rho\left(\mathbf{z}_{1:m-1}\right)\\&=\mathrm{N}\left(\mathbf{z}_m;\mathbf{h}_m,\mathbf{R}_m\right)\mathrm{N}\left(\mathbf{v}_m;\mathbf{v}_{m|m-1},\mathbf{P}_m^{-}\right)\\&\times\mathrm{IW}\left(\mathbf{P}_m^{-};\delta_m,\Delta_m\right)\mathrm{IW}\left(\mathbf{R}_m;\iota_m^{-},l_m^{-}\right)\\&\times\rho\left(\mathbf{z}_{1:m-1}\right)\end{aligned} \tag{34}$$

Bringing (34) into (32) and using the fixed-point iteration, one obtains

$$q\left(\mathbf{v}_{n,m}\right)=\mathrm{N}\left(\mathbf{v}_{n,m};\mathbf{v}_{n,m},\mathbf{P}_{n,m}\right) \tag{35}$$

$$q\left(\mathbf{P}_{n,m}^{-}\right)=\mathrm{IW}\left(\mathbf{P}_{n,m}^{-};\delta_{n,m},\Delta_{n,m}\right) \tag{36}$$

$$q\left(\mathbf{R}_{n,m}\right)=\mathrm{IW}\left(\mathbf{R}_{n,m};\hat{\iota}_{n,m},\hat{l}_{n,m}\right) \tag{37}$$

See Appendix for detailed derivation process.

Firstly, we employ (36) to compute the mean of the posterior distribution of $\mathbf{P}_{n,m}^{-}$:

$$\bar{\mathbf{P}}_{n,m}=E\left[\mathbf{P}_{n,m}^{-} \mid \hat{\mathbf{z}}_{n,1:m-1}\right]=\frac{\Delta_{n,m}}{\delta_{n,m}-\alpha-1} \tag{38}$$

where the dof parameter $\delta_m$ and the inverse scale matrix $\Delta_m$ are iteratively updated in each iteration:

$$\delta_{n,m}^{j+1}=\bar{\delta}_{n,m}+1 \tag{39}$$

$$\Delta_{n,m}^{j+1}=\bar{\Delta}_{n,m}+\mathrm{M}_{n,m}^{j+1} \tag{40}$$

$$\mathrm{M}_{n,m}^{j+1}=\frac{1}{2\alpha+1}\sum_{i=0}^{2\alpha}\left(\bar{\mathbf{v}}_{n,m}-\left(\nu_{n,m}^{i}\right)^{j+1}\right)\left(\bar{\mathbf{v}}_{n,m}-\left(\nu_{n,m}^{i}\right)^{j+1}\right)^{T} \tag{41}$$

and $\left(\nu_m^i\right)^{j+1}$ are the Sigma points generated from the results $\mathbf{v}_{n,m}^{j+1}$ and $\mathbf{P}_{n,m}^{j+1}$ obtained in the $j$+1-th iteration. Next, we can obtain an estimate of $\mathbf{R}_{n,m}$ by calculating the expectation of the IW distribution:

$$\mathbf{R}_{n,m}^{j+1}=\frac{\hat{l}_{n,m}^{j+1}}{\hat{\iota}_{n,m}^{j+1}-\beta-1} \tag{42}$$

where the dof parameter $\iota_{n,m}$ and the inverse proportionality matrix $l_{n,m}$ are obtained via:

$$\hat{\iota}_{n,m}^{j+1}=\hat{\iota}_{n,m}^{-}+1 \tag{43}$$

$$\hat{l}_{n,m}^{j+1}=\hat{l}_{n,m}^{-}+\mathbf{A}_{n,m}^{j+1} \tag{44}$$

$$\mathbf{A}_{n,m}^{j+1}=\frac{1}{2\alpha+1}\sum_{i=0}^{2\alpha}\left(\mathbf{z}_{n,m}-\mathbf{h}_{n,m}\left(\left(\nu_{n,m}^{i}\right)^{j+1}\right)\right)\left(\mathbf{z}_{n,m}-\mathbf{h}_{n,m}\left(\left(\nu_{n,m}^{i}\right)^{j+1}\right)\right)^{T} \tag{45}$$

With the updated noise covariance matrices $\bar{\mathbf{P}}_{n,m}$ and $\mathbf{R}_{n,m}$ obtained from the VB inference, we proceed to correct the state estimate using the MGST criterion. To integrate the robust kernel into the filtering framework, we construct an augmented error model based on the statistical linearization in (46). Then based on the statistical linearization method, the local measurement equation can be linearized as [7]

$$\begin{aligned}\mathbf{z}_{n,m}&=\mathbf{H}_{n,m}\left(\mathbf{v}_{n,m}-\bar{\mathbf{v}}_{n,m}\right)\\&+\mathbf{h}_{n,m}\left(\mathbf{v}_{n,m}-\bar{\mathbf{v}}_{n,m}\right)+\mathbf{r}_{n,m}+\varepsilon_{n,m}\end{aligned} \tag{46}$$

where $\mathbf{H}_{n,m}=\left(\mathbf{P}_{n,\mathbf{vz},m}^{-}\right)^{T}\left(\bar{\mathbf{P}}_{n,m}\right)^{-1}$ . And $\varepsilon_{n,m}$ is the statistical linearization error with zero mean and its covariance matrix is given by

$$\begin{aligned}\bar{\mathbf{E}}_{n,m}&=E\left[\varepsilon_{n,m}\varepsilon_{n,m}^{T}\right]\\&=\mathbf{P}_{n,\mathbf{zz},m}^{-}-\left(\mathbf{P}_{n,\mathbf{vz},m}^{-}\right)^{T}\left(\bar{\mathbf{P}}_{n,m}\right)^{-1}\mathbf{P}_{n,\mathbf{vz},m}^{-}\end{aligned} \tag{47}$$

Next, an error augmentation model is constructed to centralize the state and measurement errors for subsequent processing, the model is

$$\bar{\mathbf{z}}_{n,m}=\bar{\mathbf{H}}_{n,m}\mathbf{v}_{n,m}+\bar{\mathbf{e}}_{n,m} \tag{48}$$

where

$$\bar{\mathbf{z}}_{n,m} = \begin{bmatrix} \mathbf{v}^{-}_{n,m} \\ \mathbf{z}_{n,m} + \mathbf{H}_{n,m}\mathbf{v}^{-}_{n,m} - \hat{\mathbf{z}}^{-}_{n,m} \end{bmatrix} \in \mathbb{R}^{\alpha+\beta} \tag{49}$$

$$\bar{\mathbf{H}}_{n,m} = \begin{bmatrix} \mathbf{I} \\ \mathbf{H}_{n,m} \end{bmatrix} \in \mathbb{R}^{\alpha+\beta} \tag{50}$$

$$\bar{\mathbf{e}}_{n,m} = \begin{bmatrix} \mathbf{v}^{-}_{n,m} - \mathbf{v}_{n,m} \\ \mathbf{r}_{n,m} + \varepsilon_{n,m} \end{bmatrix} \tag{51}$$

The covariance matrix of the error $\bar{\mathbf{e}}_{n,m}$ is

$$\begin{aligned} E\left[\bar{\mathbf{e}}_{n,m}\bar{\mathbf{e}}^{T}_{n,m}\right] &= \begin{bmatrix} \mathbf{P}^{-}_{n,m} & 0 \\ 0 & \mathbf{R}_{n,m} + \bar{\mathbf{E}}_{n,m|m} \end{bmatrix} \\ &= \begin{bmatrix} \overset{P}{\mathbf{S}}_{n,m}\left(\overset{P}{\mathbf{S}}_{n,m}\right)^{T} & 0 \\ 0 & \overset{R}{\mathbf{S}}_{n,m}\left(\overset{R}{\mathbf{S}}_{n,m}\right)^{T} \end{bmatrix} \\ &= \mathbf{S}_{n,m}\left(\mathbf{S}_{n,m}\right)^{T} \end{aligned} \tag{52}$$

where $\overset{P}{\mathbf{S}}_{n,m} \in \mathbb{R}^{\alpha\times\alpha}$, $\overset{R}{\mathbf{S}}_{n,m} \in \mathbb{R}^{\beta\times\beta}$, and $\mathbf{S}_{n,m} \in \mathbb{R}^{(\alpha+\beta)\times(\alpha+\beta)}$ are obtained by Cholesky decomposition.

By left-multiplying $\mathbf{S}^{-1}_{n,m}$ on the constructed error augmentation model, we get

$$\lambda_{n,m} = \hbar_{n,m}\left(\mathbf{v}_{n,m}\right) + \ell_{n,m} \tag{53}$$

where $\lambda_{n,m} = \mathbf{S}^{-1}_{n,m}\bar{\mathbf{z}}_{n,m}$, $\hbar_{n,m} = \mathbf{S}^{-1}_{n,m}\bar{\mathbf{H}}_{n,m}$, $\ell_{n,m} = \mathbf{S}^{-1}_{n,m}\bar{\mathbf{e}}_{n,m}$.

To suppress the influence of non-Gaussian outliers contained in $\mathbf{e}_{n,m}$, we define the MGST-based cost function $J\left(\mathbf{v}_{n,m}\right)$ as the negative sum of the kernel values applied to each error component:

$$J\left(\mathbf{v}_{n,m}\right) = \sum_{i=1}^{\alpha+\beta}\left(1 + \frac{\left\|\left[\mathbf{e}_{n,m}\right]_i\right\|^{\xi}}{c\gamma^2}\right)^{-(c+\xi)/\xi} \tag{54}$$

where $\left[\mathbf{e}_{n,m}\right]_i$ denotes the *i*-th element of the augmented error vector. Minimizing (54) is equivalent to maximizing the correntropy between the estimated and true states under the generalized Student's t-kernel. To find the optimal state $\mathbf{v}_{n,m}$, we take the gradient of (54) with respect to $\mathbf{v}_{n,m}$ and set it to zero. Applying the chain rule, the derivative yields a weighted sum of error terms:

$$\begin{aligned} \frac{\partial J\left(\mathbf{v}_{n,m}\right)}{\partial \mathbf{v}_{n,m}} &= -\frac{1}{\alpha+\beta}\frac{c+\xi}{c\gamma^2}\sum_{i=1}^{\alpha+\beta}\left\|\mathbf{e}_{n,i,m}\right\|^{\xi-1}\left(1+\frac{\left\|\mathbf{e}_{n,i,m}\right\|^{\xi}}{c\gamma^2}\right)^{-(c+2\xi)/\xi} \\ &= -\frac{1}{\alpha+\beta}\sum_{i=1}^{\alpha+\beta}\underbrace{\frac{c+\xi}{c\gamma^2}\left\|\mathbf{e}_{n,i,m}\right\|^{\xi-2}\left(1+\frac{\left\|\mathbf{e}_{n,i,m}\right\|^{\xi}}{c\gamma^2}\right)^{-(c+2\xi)/\xi}}_{Weight\ \ Factor\ \ \Psi_{n,i,m}}\frac{1}{\mathbf{e}_{n,i,m}}\frac{\partial \mathbf{e}_{n,i,m}}{\partial \mathbf{v}_m} \\ &= 0 \end{aligned} \tag{55}$$

***Key Insight:*** The term in the underline acts as an adaptive weight $\Psi_{n,i,m}$. Unlike standard Gaussian kernels where the weight decay is fixed, here the decay rate is governed by $c$ while the curvature near zero is controlled by $\xi$. This allows the filter to aggressively suppress large outliers without compromising sensitivity to small residuals.

By organizing these scalar weights into a diagonal matrix $\Psi_{n,m} = diag\left(\Psi_{n,1,m}, \Psi_{n,2,m}, \ldots, \Psi_{n,\alpha+\beta,m}\right)$, equation (55) can be rewritten in matrix form. Solving for $\mathbf{v}_{n,m}$ leads to the fixed-point iteration update:

$$\mathbf{v}_{n,m} = \left(\left(\Upsilon_{n,m}\right)^{T}\Psi_{n,m}\Upsilon_{n,m}\right)^{-1}\left(\Upsilon_{n,m}\right)^{T}\Psi_{n,m}\lambda_{n,m} \tag{56}$$

Letting

$$\Psi_{n,i,m} = \left\|\mathbf{e}_{n,i,m}\right\|^{\xi-2}\left(1+\frac{\left\|\mathbf{e}_{n,i,m}\right\|^{\xi}}{c\gamma^2}\right)^{-(c+2\xi)/\xi} \tag{57}$$

$$\Psi_{n,m} = \begin{bmatrix} \Psi_{n,p,m} & 0 \\ 0 & \Psi_{n,r,m} \end{bmatrix} \tag{58}$$

$$\Psi_{n,p,m} = diag\left(\Psi_{n,1,m}, \Psi_{n,2,m}, \ldots, \Psi_{n,\alpha,m}\right) \tag{59}$$

$$\Psi_{n,r,m} = diag\left(\Psi_{n,\alpha+1,m}, \Psi_{n,\alpha+2,m}, \ldots, \Psi_{n,\alpha+\beta,m}\right) \tag{60}$$

According to the matrix inversion lemma, (56) can be rewritten as:

$$\mathbf{v}_{n,m} = \mathbf{v}^{-}_{n,m} + \mathbf{K}_{n,m}\left(\mathbf{z}_{n,m} - \hat{\mathbf{z}}^{-}_{n,m}\right) \tag{61}$$

$$\mathbf{K}_{n,m} = \bar{\mathbf{P}}^{-}_{n,m}\mathbf{H}_{n,m}\left(\mathbf{H}_{n,m}\bar{\mathbf{P}}^{-}_{n,m}\mathbf{H}^{T}_{n,m} + \bar{\mathbf{R}}_{n,m}\right)^{-1} \tag{62}$$

$$\bar{\mathbf{P}}^{-}_{n,m} = \overset{P}{\mathbf{S}}_{n,m}\Psi_{n,p,m}\left(\overset{P}{\mathbf{S}}_{n,m}\right)^{T} \tag{63}$$

$$\bar{\mathbf{R}}_{n,m} = \overset{R}{\mathbf{S}}_{n,m}\Psi_{n,r,m}\left(\overset{R}{\mathbf{S}}_{n,m}\right)^{T} \tag{64}$$

Finally, the error covariance matrix of the algorithm is updated by:

$$\begin{aligned} \mathbf{P}_{n,m} &= \left(\mathbf{I} - \mathbf{K}_{n,m}\mathbf{H}_{n,m}\right)\mathbf{P}^{-}_{n,m}\left(\mathbf{I} - \mathbf{K}_{n,m}\mathbf{H}_{n,m}\right)^{T} \\ &+ \mathbf{K}_{n,m}\left(\mathbf{R}_{n,m} + \bar{\mathbf{E}}_{n,m}\right)\mathbf{K}^{T}_{n,m} \end{aligned} \tag{65}$$

For the subsequent design of the discrete state estimator, we transform the local state estimation results and the local error covariance matrix into informative form.

$$\chi_{n,m} = \mathbf{C}_{n,m}\mathbf{v}_{n,m} \tag{66}$$

$$\mathbf{C}_{n,m} = \left(\mathbf{P}_{n,m}\right)^{-1} \tag{67}$$

where $\chi_{n,m}$ and $\mathbf{C}_{n,m}$ represent the information vector and information matrix, respectively.

This structure explicitly separates the robust state correction (via MGST) from the statistical noise learning (via VB), improving numerical stability compared to joint optimization methods.

### *C. Decentralized MGST Variational Bayesian UKF*

In this subsection, we give the inter-region fusion algorithm using the maximum posteriori state estimation method to update the local state estimation results using the edge measurement information. For $n$-th region, the updated state estimation results and the estimation error covariance matrix are as follows:

$$\mathbf{v}_{n,m}=\mathbf{P}_{n,m}\left(\chi_{n,m}+\sum_{i\in N_n}\mathbf{H}_{ni,m}^{T}\overline{\mathbf{E}}_{ni,m}^{-1}\tilde{\mathbf{z}}_{ni,m}\right) \tag{68}$$

$$\mathbf{P}_{n,m}=\left(\mathbf{C}_{m}+\sum_{i\in N_n}\mathbf{H}_{ni,m}^{T}\overline{\mathbf{E}}_{ni,m}^{-1}\mathbf{H}_{ni,m}\right)^{-1} \tag{69}$$

where

$$\mathbf{H}_{ni,m}=\left(\mathbf{P}_{ni,\mathbf{vz},m}^{-}\right)^{T}\mathbf{P}_{n,m|m-1}^{-1} \tag{70}$$

$$\overline{\mathbf{E}}_{ni,m}=\mathbf{E}_{ni,m}+\mathbf{H}_{ni,m}\mathbf{C}_{i,m}^{-1}\mathbf{H}_{ni,m}^{T} \tag{71}$$

---

Algorithm 1: D-MGST-VBUKF

---

**For** $n=1,2,\ldots,N$ do

**1: Input:** $\mathbf{f}(\bullet)$, $\mathbf{h}_{n,m}(\bullet)$, $c$, $\gamma$, $\xi$, $N$, $\lambda$, $\eta$, $\tau$, $\varsigma$, $\zeta$.

**2: Output:** $\mathbf{v}_{n,m}$ for $m=1,2,\ldots,M$

**3: Intitialization:** Setting initial filter state values $\mathbf{v}_{n,0}$, initial covariance matrix values $\mathbf{P}_{n,0}$, $\Delta_{n,0}$ and $\hat{l}_{n,0}$.

**4: for** $m=1,2,\ldots,M$ **do**

The priori state estimates $\mathbf{v}_{n,m}^{-}$ and the priori covariance matrix $\mathbf{P}_{n,m}^{-}$ are computed by (13)-(16). And the priori Bayesian parameters $\delta_{n,m}^{-}$, $\Delta_{n,m}^{-}$, $\hat{\iota}_{n,m}^{-}$ and $\hat{l}_{n,m}^{-}$ can be computed by (17)-(25).

**for** $j=1,2,\ldots,J$

Update the posterior Bayesian parameters $\delta_{n,m}^{j+1}$, $\Delta_{n,m}^{j+1}$, $\hat{\iota}_{n,m}^{j+1}$ and $\hat{l}_{n,m}^{j+1}$ via (30)-(45). Then compute the estimation error covariance matrix $\mathbf{P}_{n,m}^{-}$ and noise covariance matrix $\mathbf{R}_{n,m}^{j+1}$ by (38) and (42), respectively.

Compute the $\mathbf{v}_{n,m}$ by (48)-(61).

**end for**

Updating the covariance matrix $\mathbf{P}_{n,m}$ by (65).

Compute $\chi_{n,m}$ and $\mathbf{C}_{n,m}$ by (66) and (67), respectively.

The final state estimation results $\mathbf{v}_{n,m}$ and error covariance matrix $\mathbf{P}_{n,m}$ are obtained by (68)-(74) fusing the neighboring region information.

**end for**

**End For**

---

$$\tilde{\mathbf{z}}_{ni,m}=\tilde{\mathbf{z}}_{ni,m}-h_{ni}\left(\mathbf{v}_{n,m}^{-},\mathbf{v}_{i,m}^{-}\right)+\mathbf{H}_{ni,m}\mathbf{v}_{n,m}^{-}+\mathbf{H}_{in,m}\mathbf{v}_{i,m}^{-}-\mathbf{H}_{in,m}\chi_{i,m}^{-1}\mathbf{C}_{i,m} \tag{72}$$

with

$$\mathbf{P}_{ni,\mathbf{vz},m}^{-}=\sum_{j=0}^{2\alpha}\omega_c^j\left[\nu_{n,m|m-1}^{j}-\mathbf{v}_{n,m}^{-}\right]\left[h_{ni}\left(\nu_{n,m|m-1}^{j},\mathbf{v}_{i,m}^{-}\right)-\hat{\mathbf{z}}_{ni,m}^{-}\right]^{T} \tag{73}$$

$$\hat{\mathbf{z}}_{ni,m}^{-}=\sum_{j=0}^{2\alpha}\omega_m^i h_{ni}\left(\nu_{n,m|m-1}^{j},\mathbf{v}_{i,m}^{-}\right) \tag{74}$$

This decentralized architecture enhances scalability and resilience by eliminating the need for a central fusion node and only requiring the exchange of processed residuals at boundaries, preserving data privacy. The complete procedure for the proposed D-MGST-VBUKF is summarized in Algorithm 1.

***Remark 5:*** Comparison with Prior Art:

vs. RD-UKF [7]: The RD-UKF employs robust loss functions but requires predefined noise covariance matrices $\mathbf{Q}_{n,m}$ and $\mathbf{R}_{n,m}$. In contrast, our method online learns these matrices via VB, making it suitable for unknown noise environments.

vs. Standard VBKF [31]: Traditional VBKF assumes Gaussian likelihoods, which degrade under heavy-tailed noise. Our integration of the MGST kernel within the VB framework preserves robustness against outliers while retaining the adaptive noise estimation capability.

vs. Centralized MGST: By decentralizing the computation and fusing only boundary information (Eq. 68-74), we reduce communication bandwidth by approximately O($N$) compared to centralized implementations, where $N$ is the number of regions.

## V. SIMULATION

In this section, the performance of the proposed D-MGST-VBUKF algorithm is verified in an IEEE 14-bus system and IEEE 39-bus system. In order to fully demonstrate the technical advantages of the method, the traditional centralized UKF, PF [32], HIF [33], CKMC-UKF [14] and the RD-UKF [7] are selected as comparison benchmarks for the experiments. The simulations are done in MATLAB R2020a platform and Matpower toolbox [34] environment, and all test cases are executed on PCs with Intel Core i7-11700HQ processors. The model parameters of the IEEE-14 bus system are set to $\mathbf{F}_k=0.89\mathbf{I}_2$, $\mathbf{G}_k=0.11\mathbf{I}_2$, and $\mathbf{Q}_k=0.01^2\mathbf{I}_2$. The desired steady state is given in the Matpower toolbox [34]. The position settings of SCADA unit and PMU are shown in Fig. 1. The IEEE14 bus system area is divided into three areas, and the IEEE39 bus system area is divided into four areas as shown in Fig. 1, with relevant measurement information labeled. The unscented transformation parameters used in all regions are set to $\lambda=e^2$, $\eta=0.02$ and $\tau=1$. In addition, the parameter settings of the RD-UKF used to compare the algorithms are the same as in [7]. In this paper, the root-mean-square error (RMSE) and average RMSE (ARMSE) are used as the criterion for evaluating the performance of the algorithms, which is given by

$$\mathrm{RMSE}=\sqrt{\frac{1}{L}\sum_{i=1}^{L}\left(\mathbf{v}_{i,m}-\mathbf{v}_{i,m}\right)^{2}} \tag{75}$$

$$\mathrm{ARMSE}=\frac{1}{M}\sqrt{\frac{1}{L}\sum_{i=1}^{L}\left(\mathbf{v}_{i,m}-\mathbf{v}_{i,m}\right)^{2}} \tag{76}$$

where $L$ stands for represents the number of Monte Carlo experiments and $M$=100 represents the total time step. All the experiments in this paper were conducted with $L$=100 Monte Carlo experiments to ensure statistical properties.

Table II
ARMSE (°) of Voltage phase of different parameter at Bus 1 in IEEE 14-bus system

| Parameter | $\xi$=1.8 | $\xi$=1.9 | $\xi$=2.0 | $\xi$=2.1 | $\xi$=2.2 |
|---|---|---|---|---|---|
| $\gamma$=8 | 0.417794 | 0.345325 | 0.396547 | 0.860261 | 0.447836 |
| $\gamma$=6 | 0.444927 | 0.345158 | 0.356240 | 0.618626 | 0.484056 |
| $\gamma$=10 | 0.425725 | 0.336041 | 0.425067 | 0.981710 | 0.493159 |
| $\gamma$=12 | 0.400874 | **0.330987** | 0.457528 | 0.985102 | 0.479071 |
| $\gamma$=14 | 0.392918 | 0.358319 | 0.471917 | 1.024134 | 0.495052 |

Table III
ARMSE (V) of Voltage amplitude of different parameter at Bus 1 in IEEE 14-bus system

| Parameter | $\xi$=1.8 | $\xi$=1.9 | $\xi$=2.0 | $\xi$=2.1 | $\xi$=2.2 |
|---|---|---|---|---|---|
| $\gamma$=8 | 0.333967 | **0.144329** | 0.228805 | 1.350270 | 0.360520 |
| $\gamma$=6 | 0.353814 | 0.171393 | 0.187110 | 0.973702 | 0.396835 |
| $\gamma$=10 | 0.310811 | 0.168024 | 0.322359 | 1.451856 | 0.414089 |
| $\gamma$=12 | 0.290297 | 0.181415 | 0.361682 | 1.015709 | 0.383568 |
| $\gamma$=14 | 0.295684 | 0.218233 | 0.388386 | 1.450091 | 0.391932 |

### *A. Parameter Sensitivity and Tuning Analysis*

The MGST framework proposed in this paper introduces new shape parameters designed to enhance the algorithm's robustness in complex noisy environments. To this end, this section systematically examines the parameter settings to evaluate the impact of these newly introduced parameters on algorithmic performance. Set $c=2$ and change the other two parameters for testing. The experiments are conducted under a mixed Gaussian noise environment, with its distribution given as follows:

$$\mathbf{r}_{n,m}=0.01N\left(0,100\right)+0.99N\left(0,1\right) \tag{77}$$

As shown in Tables II and III, which present the ARMSE of the algorithm under different parameters in the noisy environment, $\xi$=1.9, $\gamma$=12 and $\xi$=1.9, $\gamma$=8 exhibit the minimum ARMSE values for amplitude and phase angle, respectively, indicating that the algorithm performs optimally under these conditions. Furthermore, it can be observed that variations in parameter $\gamma$ do not significantly affect the algorithm’s performance, whereas parameter $\xi$ is the primary factor influencing performance. Therefore, the newly introduced parameters play a crucial role in enhancing the performance of the MGST algorithm when dealing with complex noise environments. In the subsequent simulation, the parameter combination set is $c=2, \xi$=1.9, $\gamma$=12.

### *B. Gaussian noise environment*

In this setting, we assume that the noise interference to the measurements is Gaussian distributed. For the UKF as well as the RD-UKF algorithms it is necessary to set the measurement noise covariance matrix, which we set to $\mathbf{R}_{n,m}=1\times10^{-3}$ and $\mathbf{R}_{nj,m}=1\times10^{-3}$ ( $n, j=\left(1,2,\ldots,14\right), n\neq j$ ). To evaluate the filtering performance of the D-MGST-VBUKF algorithm proposed in this study in complex noise environments, we show the results of the RMSE comparison experiments for bus in Figs. 3-6. The experimental results show that the D-MGST-VBUKF algorithm exhibits excellent robustness under strong Gaussian noise interference scenarios, and its voltage magnitude and phase angle estimation trajectories are able to closely track the real state. In contrast, the RD-UKF method shows phase angle tracking deviation during the dynamic process, which verifies that the algorithm in this paper effectively improves the state estimation accuracy of the power system through the synergistic optimization based on the MGST and the VB framework.

### *C. Non-Gaussian noise environment*

In view of the prevalent non-Gaussian noise interference problem in power system state estimation, in order to verify the adaptability of this algorithm in complex non-Gaussian environments, this subsection constructs a mixed noise scenario with typical non-Gaussian characteristics for verification experiments. According to the statistical characteristics of the actual power system noise, this study adopts the superposition model of the hybrid Gaussian distribution and the heavy-tailed Laplace distribution, and constructs two types of typical non-Gaussian noise models respectively: (1) the hybrid Gaussian noise model, which simulates the multimodal noise distribution by setting up multiple Gaussian components of the differentiated mean and variance; and (2) the Laplace-Gaussian hybrid noise model, which makes use of the spiking characteristics of the Laplace distribution to simulate impulse noise disturbances. This composite noise modeling method can effectively characterize the statistical properties of complex disturbances such as measurement anomalies and data loss in the power system, and provide a rigorous test benchmark for algorithm

Table IV
ARMSE (°) of Voltage phase of different algorithm at different Bus in IEEE 14-bus under noise $\mathbf{r}1_{n,m}$

| | UKF | PF | HIF | CKMC-UKF | RD-UKF | D-MGST-VBUKF |
|---|---|---|---|---|---|---|
| Bus 1 | 0.277396 | 0.543141 | 0.617627 | 0.380861 | 0.430993 | **0.151320** |
| Bus 3 | 0.972165 | 0.563662 | 0.590819 | 0.730539 | 0.163731 | **0.160412** |
| Bus 6 | 1.116158 | 0.389168 | 0.628008 | 0.774075 | 0.196619 | **0.174360** |
| Bus 11 | 1.235606 | 0.441383 | 0.609957 | 0.763933 | 0.201352 | **0.157360** |
| Bus 14 | 0.833729 | 0.570619 | 0.615180 | 1.365522 | 0.165903 | **0.159579** |

Table V
ARMSE (V) of Voltage amplitude of different algorithm at different Bus in IEEE 14-bus under noise $\mathbf{r}1_{n,m}$

| | UKF | PF | HIF | CKMC-UKF | RD-UKF | D-MGST-VBUKF |
|---|---|---|---|---|---|---|
| Bus 1 | 0.282945 | 0.381197 | 0.597902 | 0.392859 | 0.461407 | **0.262792** |
| Bus 3 | 0.537960 | 0.766054 | 0.759203 | 0.697780 | 0.536580 | **0.386193** |
| Bus 6 | 2.662729 | 0.760417 | 0.725732 | 0.880946 | 0.562215 | **0.350157** |
| Bus 11 | 0.417463 | 0.784607 | 0.739559 | 0.888215 | 0.495963 | **0.323091** |
| Bus 14 | 0.460778 | 0.798523 | 0.819593 | 1.574965 | 0.590152 | **0.416465** |

Table VI
ARMSE (°) of Voltage phase of different algorithm at different Bus in IEEE 39-bus under noise $\mathbf{r}2_{n,m}$

| | UKF | PF | HIF | CKMC-UKF | RD-UKF | D-MGST-VBUKF |
|---|---|---|---|---|---|---|
| Bus 1 | 13.155345 | 0.411432 | 0.614891 | 0.503000 | 0.745586 | **0.289302** |
| Bus 7 | 173.628383 | 0.384853 | 0.599933 | 1.132608 | 0.680222 | **0.367549** |
| Bus 18 | 2.804112 | 0.374730 | 0.565873 | 0.621632 | 0.623504 | **0.318918** |
| Bus 28 | 3.040500 | 0.444437 | 0.625420 | 0.701509 | 5.183148 | **0.122148** |
| Bus 32 | 60.914009 | 0.626693 | 0.563290 | 2.258223 | 0.527844 | **0.279371** |

Table VII
ARMSE (V) of Voltage amplitude of different algorithm at different Bus in IEEE 39-bus under noise $\mathbf{r}2_{n,m}$

| | UKF | PF | HIF | CKMC-UKF | RD-UKF | D-MGST-VBUKF |
|---|---|---|---|---|---|---|
| Bus 1 | 6.788971 | 0.844352 | 0.833884 | 0.775040 | 0.856605 | **0.473376** |
| Bus 7 | 98.251337 | 0.756075 | 0.751982 | **0.568663** | 0.757396 | 0.742935 |
| Bus 18 | 0.390663 | 0.707115 | 0.689162 | 0.890833 | 0.668142 | **0.378749** |
| Bus 28 | 1.725703 | 0.613166 | 0.662923 | 0.986960 | 1.786398 | **0.196507** |
| Bus 32 | 7.382806 | 0.658140 | 0.643215 | 1.938584 | 0.665735 | **0.501140** |

robustness evaluation. The following two noise distributions are modeled

$$\mathbf{r}1_{n,m} = 0.01N(0,1000) + 0.99N(0,1) \tag{78}$$

$$\mathbf{r}2_{n,m} = 0.01L(0,1000) + 0.99N(0,1) \tag{79}$$

where $N(a,b)$ is the Gaussian distribution with mean $a$ and variance $b$, and $L(a,b)$ is the Laplace distribution with mean $a$ and variance $b$. The ARMSE comparison results shown in Table IV-VII indicate that the innovative algorithm proposed in this study achieves a significant improvement in state estimation accuracy on multiple test buses under the same experimental conditions. Compared to the UKF, PF, HIF, CKMC-UKF and RD-UKF methods that need to rely on the Matlab's $\mathrm{Var}(\bullet)$ function to compute the measurement noise covariance offline, the present algorithm accomplishes the online estimation of the noise statistical properties synchronously during the iteration process by embedding an adaptive estimation mechanism, which avoids the need for explicit inputs of the $\mathbf{Q}$ and the $\mathbf{R}$. This unique adaptive feature not only reduces the dependence of the algorithm on the a priori knowledge of the model, but more importantly, it fundamentally solves the problem of estimation bias accumulation due to the mismatch of noise parameters in the traditional method, which provides a theoretical guarantee to cope with the time-varying noise characteristics in the dynamic process of the power system.

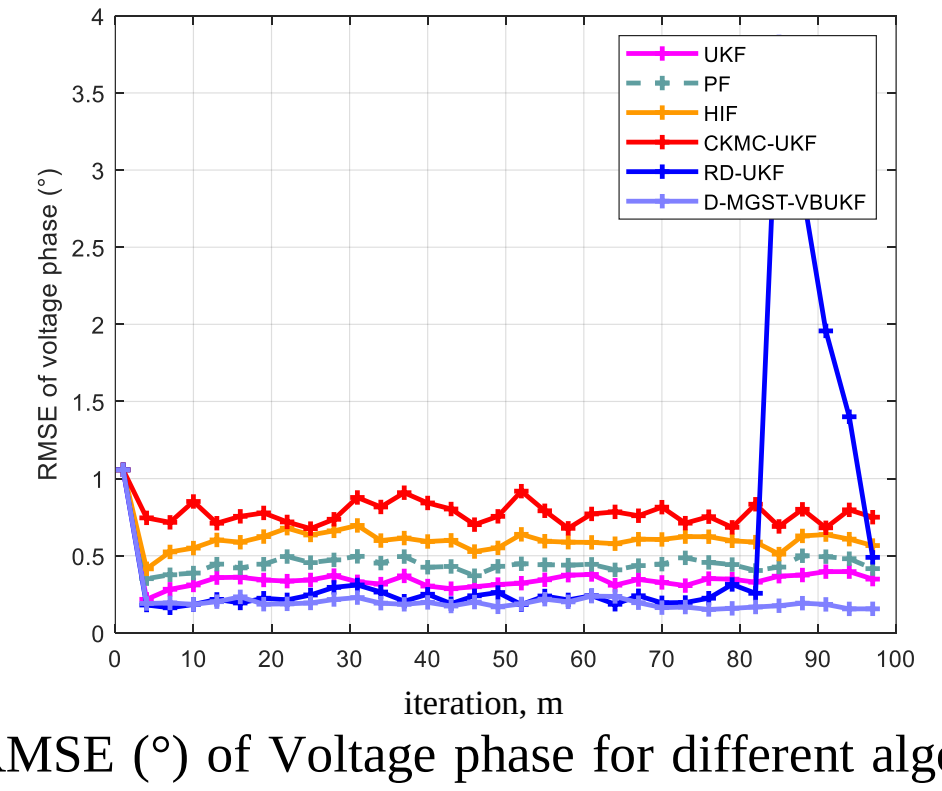

Fig. 3. RMSE (°) of Voltage phase for different algorithms at Bus 11 at IEEE 14-bus.

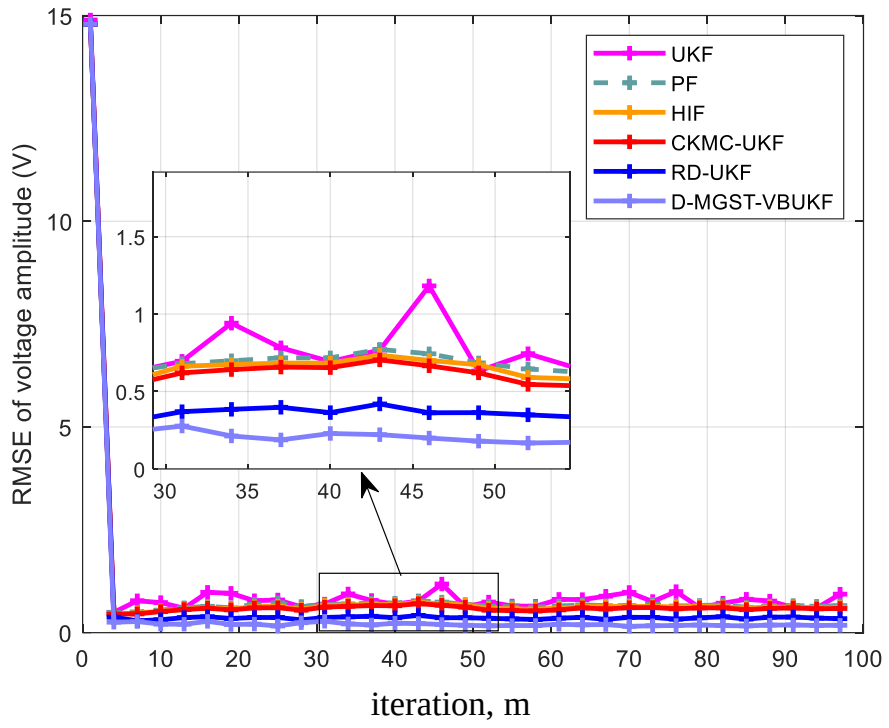

Fig. 4. RMSE (V) of Voltage amplitude for different algorithms at Bus 11 at IEEE 14-bus.

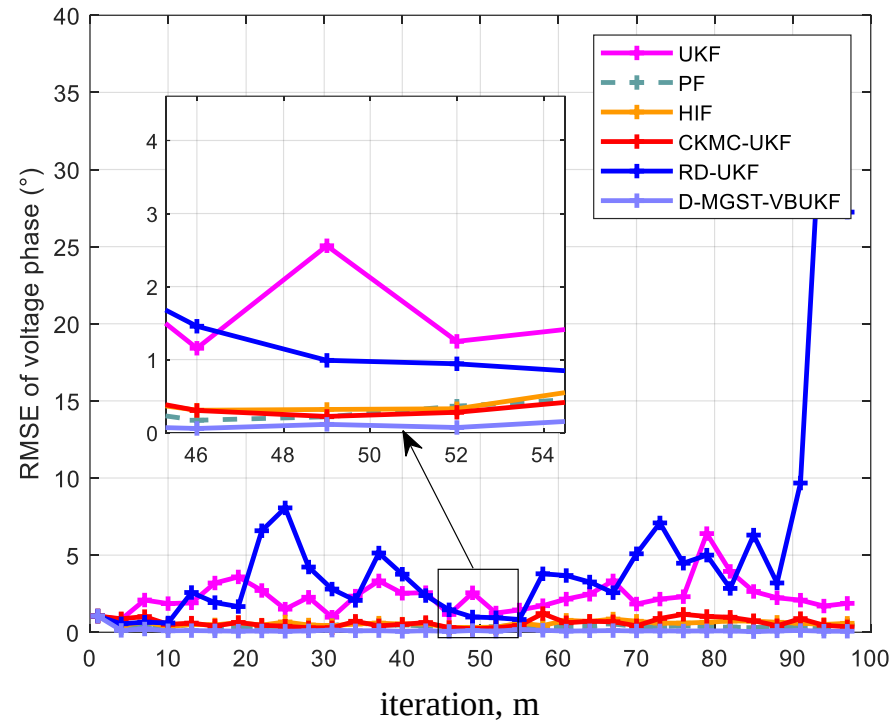

Fig. 5. RMSE (°) of Voltage phase for different algorithms at Bus 26 at IEEE 39-bus.

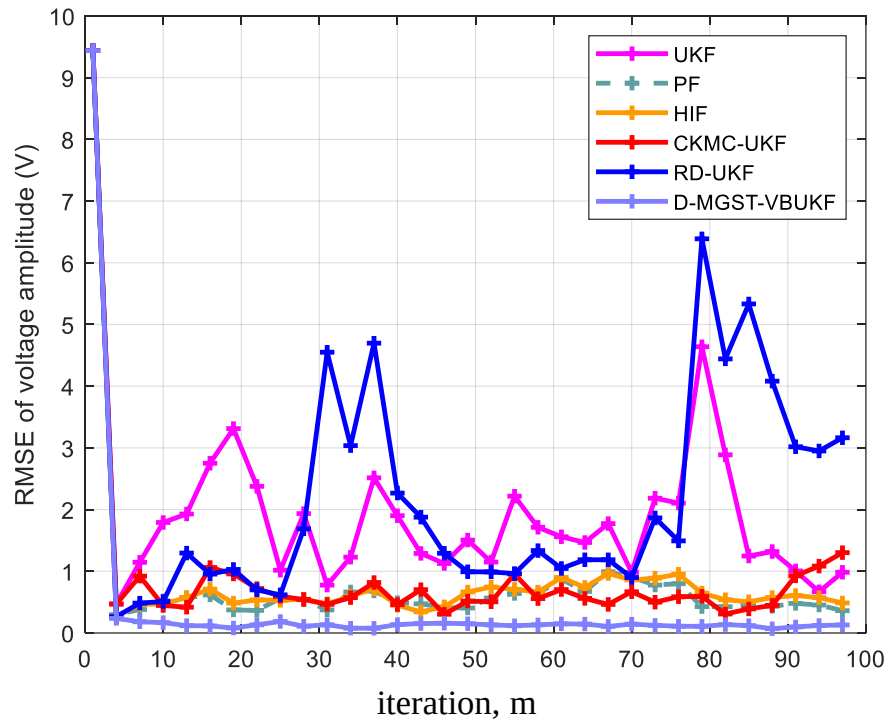

Fig. 6. RMSE (V) of Voltage amplitude for different algorithms at Bus 26 at IEEE 39-bus.

### *D. Abnormal measurement data*

In order to verify the fault-tolerant performance of the algorithm in the case of abnormal measurement data, a sudden data perturbation experiment with practical engineering significance is designed in this study. An 75% amplitude decay perturbation (i.e., measurements multiplied by 0.75) is applied to all measurements in region 1 at the $m$=55 sampling moment when the system is in stable operation, an operation that is equivalent to simulating anomalous operating conditions typical of power systems, such as transient sensor failures or communication anomalies. With the hybrid Gaussian noise parameters set in Scheme A kept constant and the algorithm hyperparameters exactly the same as in Scheme A, Fig. 7-Fig. 8 compare the RMSE of different buses in IEEE 39-bus system under this perturbation, and it can be seen that all algorithms are subject to a certain amount of fluctuation in the 55-th moment, but our proposed D-MGST-VBUKF algorithm receives the smallest disturbance, and recovers the speed is the fastest.

### *E. Computing time*

To evaluate the computational efficiency of the algorithm in practice, we measure its single-iteration runtime in Table VIII. It can observed that the proposed D-MGST-VBUKF incurs a relatively high computational cost, which is primarily due to the incorporation of variational Bayesian inference and the MGST kernel. Nevertheless, the increase in computational burden is within an acceptable range, especially considering the significant improvement in robustness.

Table VIII
Computing time (s) of different algorithm in IEEE 14-bus and IEEE 39 bus

| | IEEE 14-bus | IEEE 39-bus |
|---|---|---|
| UKF | 0.00615 | 0.02462 |
| PF | 0.00947 | 0.26023 |
| HIF | 0.00983 | 0.04070 |
| CKMC-UKF | 0.01521 | 0.02826 |
| RD-UKF | 0.03003 | 0.04375 |
| D-MGST-VBUKF | 0.03284 | 0.14290 |

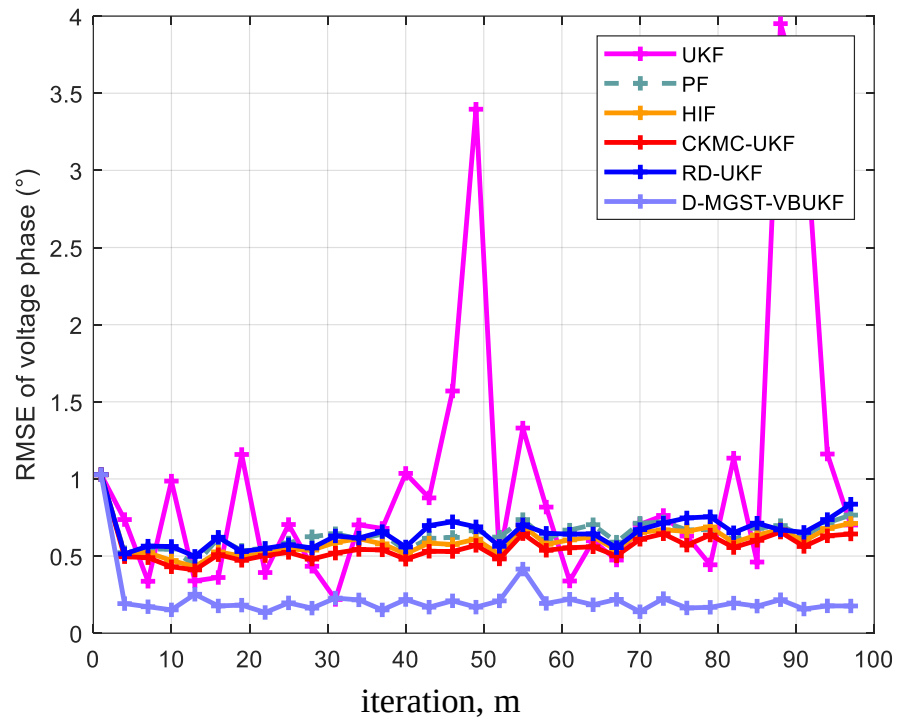

Fig. 7. RMSE (°) of Voltage phase for different algorithms at Bus 37 at IEEE 39-bus.

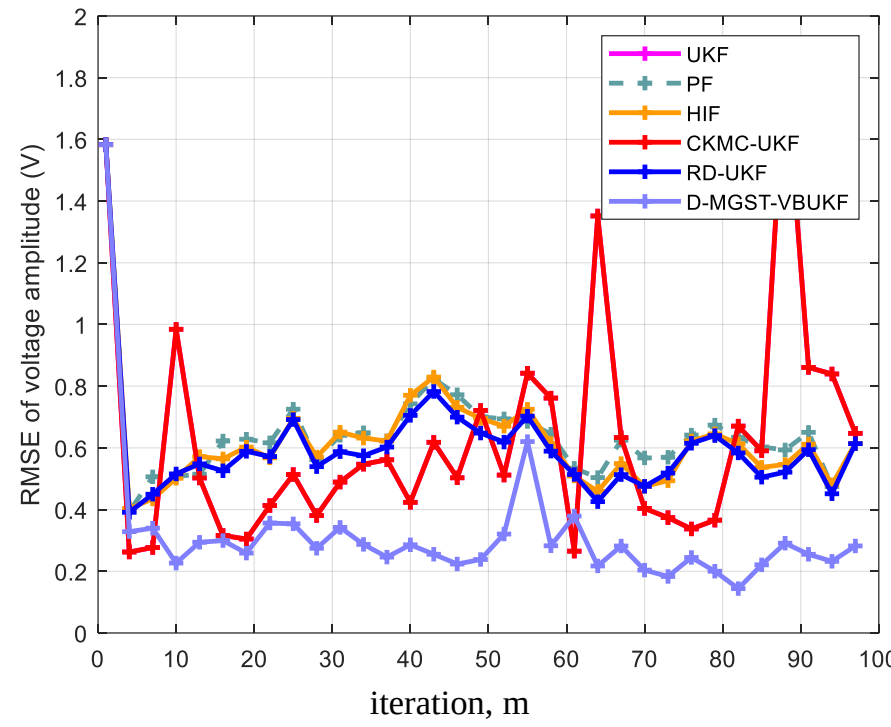

Fig. 8. RMSE (V) of Voltage amplitude for different algorithms at Bus 37 at IEEE 39-bus.

## V. Conclusion

This paper proposes a decentralized maximum generalized Student's t-kernel correntropy variational bayesian UKF (D-

MGST-VBUKF) for wide-area power systems. The algorithm is designed to address the problem of non-Gaussian noise and the unknown noise problem in WA-PSSE by the MGST and the robust variational Bayesian framework, respectively. And the decentralized method is used to fuse the information of different regions based on the interaction of buses between regions. Through comparative experiments, it is concluded that the method proposed in this paper has stronger robustness compared with the existing algorithms and can effectively face various complex situations in reality.

## APPENDIX

Take $\varphi = \mathbf{v}_{n,m}$. Then $\Theta_m^- = \left\{ \mathbf{P}_{n,m}^-, \mathbf{R}_m \right\}$. Substituting the joint distribution into (32) and retaining only terms that depend on $\mathbf{v}_{n,m}$ yields

$$\begin{aligned} \log q\left(\mathbf{v}_{n,m}\right) &= \mathrm{E}_{\mathbf{P}_{n,m}^-, \mathbf{R}_m}\left[ \log \mathrm{N}\left(\mathbf{z}_{n,m}; \mathbf{h}_{n,m}\left(\mathbf{v}_{n,m}\right), \mathbf{R}_{n,m}\right)\right] \\ &+ \log \mathrm{N}\left(\mathbf{v}_{n,m}; \mathbf{v}_{n,m|m-1}, \mathbf{P}_m^-\right) + \text{const} \end{aligned} \tag{80}$$

Writing the log-Gaussians explicitly

$$\begin{aligned} \log \mathrm{N}\left(\mathbf{z}_{n,m}; \mathbf{h}_{n,m}\left(\mathbf{v}_{n,m}\right), \mathbf{R}_{n,m}\right) &= -\frac{1}{2}\left(\mathbf{z}_{n,m} - \mathbf{h}_{n,m}\left(\mathbf{v}_{n,m}\right)\right)^T \mathbf{R}_{n,m}^{-1} \\ &\times\left(\mathbf{z}_{n,m} - \mathbf{h}_{n,m}\left(\mathbf{v}_{n,m}\right)\right) + \text{const} \end{aligned} \tag{81}$$

$$\begin{aligned} \log \mathrm{N}\left(\mathbf{v}_{n,m}; \mathbf{v}_{n,m}^-, \mathbf{P}_{n,m}^-\right) &= -\frac{1}{2}\left(\mathbf{v}_{n,m} - \mathbf{v}_{n,m}^-\right)^T \left(\mathbf{P}_{n,m}^-\right)^{-1} \\ &\times\left(\mathbf{v}_{n,m} - \mathbf{v}_{n,m}^-\right) + \text{const} \end{aligned} \tag{82}$$

Taking expectations with respect to $q\left(\mathbf{P}_{n,m}^-\right)$ and $q\left(\mathbf{R}_m\right)$ replaces the random matrices by their current mean estimates:

$$\overline{\left(\mathbf{P}_{n,m}^-\right)}^{-1} = \mathrm{E}_{q\left(\mathbf{P}_{n,m}^-\right)}\left[\left(\mathbf{P}_{n,m}^-\right)^{-1}\right]; \overline{\mathbf{R}_{n,m}}^{-1} = \mathrm{E}_{q\left(\mathbf{R}_{n,m}\right)}\left[\mathbf{R}_{n,m}^{-1}\right] \tag{83}$$

For an inverse Wishart distributed matrix $\mathrm{X} \sim \mathrm{IW}\left(\nu, \Psi\right)$, the expectation of its inverse is $\mathrm{E}\left[\mathrm{X}^{-1}\right] = \nu, \Psi^{-1}$. In our iterative scheme these expectations are computed using the most recent parameter estimates. Consequently, $q\left(\mathbf{v}_{n,m}\right)$ becomes a quadratic form in $\mathbf{v}_{n,m}$, which implies that $\mathbf{v}_{n,m}$ is Gaussian:

$$q\left(\mathbf{v}_{n,m}\right) = \mathrm{N}\left(\mathbf{v}_{n,m}; \mathbf{v}_{n,m}, \mathbf{P}_{n,m}\right) \tag{84}$$

where the mean $\mathbf{v}_{n,m}$ and covariance $\mathbf{P}_{n,m}$ are obtained by solving the corresponding linearized system—this is exactly the expression given in (35). The derivation of (36) and (37) is the same as this process, and will not be repeated here.